\documentclass[12pt]{article}
\usepackage[utf8]{inputenc}
\usepackage{float}
\usepackage{amsmath}
\usepackage{amssymb}
\usepackage{esvect}
\usepackage[letterpaper, margin=0.7 in]{geometry}
\usepackage{authblk}
  \usepackage{amsmath}
    \usepackage{amssymb}
    \usepackage{amsthm}
    \usepackage{amsfonts}
    \usepackage{braket}
\usepackage{amsmath}
\usepackage{amsfonts}
\usepackage{amssymb}
\usepackage{hyperref}
\usepackage{graphicx}
\usepackage{caption}
\usepackage{subcaption}
\usepackage{float}
\usepackage{mathtools}
\usepackage{slashed}
\usepackage{amsfonts}

\small
\title{Duality between  the massive  sine-Gordon and the massive Schwinger models at finite temperature
\\}
\author{Mir Mehedi Faruk$^{1}$\thanks{Corresponding author:  mir.faruk@mail.mcgill.ca},
} 

\affil{
McGill University
Montreal, QC H3A 2T8, Canada$^1$\\
}

\begin{document}
\small
\maketitle
\begin{abstract} 
 The  massive Schwinger and the massive sine-Gordon models are proved to be equivalent at finite temperature,
 using the path-integral framework.  The well known relations among the parameters of these models to establish the duality at $T=0$, also remain valid at 
non zero temperature.
\end{abstract}
\section{Introduction}
Dualities in 1+1
dimensional
quantum field theories are very well explored\cite{old} since Coleman first reported the duality between 
the sine-Gordon and the Thirring model \cite{cole}. Under the duality map, the soliton of the sine-Gordon theory is mapped  to the fundamental fermion of the Thirring model
and the meson states of the SG theory to the fermion anti-fermion bound states. This duality has been extended for the finite temperature case  using the path integral approach developed by Naon and others\cite{naonprd, chupa, naon, others}. 
Likewise, 
there is also 
a duality in  between the massive Schwinger model and the  massive sine-Gordon model\cite{naonprd} but the finite temperature extension of this duality
has not been reported yet.
It would be intriguing to 
see if 
the compactification
of the time variable into a circle of radius $\beta=1/T$ preserves the
equivalence between the massive sine-Gordon and the massive Schwinger models  at a fixed radius $\beta$ (or at a  fixed temperature).
\\ \\
The path integral approach to study 1+1 dimensional field theories developed by Naon and others has been very useful to understand dualities\cite{naonprd,chupa}. Note, neither the massive Schwinger nor the massive Thirring models are exactly solvable.
 Both of them can be solved using this path integral method
by treating the massless part exactly while
 doing a perturbative expansion over the mass parameter\footnote{
 Massive/massless sine-Gordon models are also not exactly solvable. Same treatment is also 
 done in this side where we treat the potential term of the sine-Gordon perturbatively.}\cite{naonprd}.
 The partition function of the massive Thirring (massive Schwinger) is then identified term by term with the sine-Gordon model (massive sine-Gordon) with appropriate duality transformation first suggested by Coleman\cite{cole}. 
 The finite temperature extension of such a duality has been successfully reported for the massive Thirring/sine-Gordon\cite{chupa} but not yet for the massive  Schwinger/massive sine-Gordon. One of the main reasons why it is not yet reported is due to the confusion with the finite temperature spectrum of
 exactly solvable sector (masless Schwinger model). 
 According to the path integral approach of M. V. Manias, C. M. Naon and M. L. Trobo\cite{naon},
the thermodynamic partition function of the  massless Schwinger model consists of a massive boson, a massless fermion along with
 zero-mass gauge excitation. But the study of 
 S. T. Love\cite{love} in the operator approach concludes 
 that this same 
 theory is completely  equivalent to an ensemble of noninteracting, neutral, massive, Bose particles of mass $\frac{e}{\sqrt{\pi}}$.
This discrepancy
can be taken care of by following some very simple steps (see Appendix) and 
it is found that
the 
 finite temperature particle spectrum of this sector 
 is nothing but a massive free boson of mass $\frac{e}{\sqrt{\pi}}$. Therefore,
 the duality between massive Schwinger and massive sine-Gordon
models can hopefully be extended using the path integral approach\cite{chupa, naonprd,naon} at finite temperature background.
That  would be the main goal of this paper.
 \\ \\
\section{The massive Schwinger and
massive sine-Gordon models
at finite temperature
}
We start with the 
partition function of the massive Schwinger model 
in the imaginary time formalism using  the path integral
approach at finite temperature\cite{1,2}. The Lagrangian density
of the well-known massive Schwinger model,
\begin{eqnarray}
\mathcal{L}_{S}=-\bar{\psi} (i\slashed{\partial}+eA)\psi +im_0\bar{\psi}\psi -\frac{1}{4}F^{\mu \nu}F_{\mu\nu}, 
\end{eqnarray}
where,
\begin{eqnarray}
\gamma_0=
\begin{bmatrix}
    0  &  1      \\
    1  &  0      
\end{bmatrix},
\gamma_1= 
\begin{bmatrix}
    0  &  i     \\
    -i  &  0      
\end{bmatrix},
\end{eqnarray}
The partition function in
Euclidean (1+1)-dimensional space-time is given,
\begin{eqnarray}
Z_S=N_0 N_\beta\int \mathcal{D}A_\mu \mathcal{D}\psi\mathcal{D}\bar{\psi} e^{-\int d^2x \mathcal{L}_s},
\end{eqnarray}
where $\int d^2 x =\int_{0}^{\beta}  dx_0 \int_{-\infty}^{\infty} dx_1$ with $\beta=\frac{1}{T}$.
Here $N_0$ denotes an infinite constant which does
not depend on temperature
whereas $N_\beta$ is a
temperature-dependent constant which is to be determined from the free partition
function\footnote{see for instance eq. 2.2 of ref. \cite{naon}}.
The functional integral is carried over fermionic fields obeying
antiperiodic boundary conditions in the time direction, 
\begin{eqnarray}
\psi(x_0,x_1)=-\psi(x_0+\beta,x_1).
\end{eqnarray}
Now doing the decoupling transformation,
\begin{eqnarray}
&&\psi=e^{\gamma_5 \phi}\chi,\\
&&\bar{\psi}=\bar{\chi}e^{\gamma_5 \phi},\\
&&A_\mu= -\frac{1}{e}\epsilon_{\mu\nu}\partial_\nu \phi + \partial_\mu \eta.
\end{eqnarray}
Here
the bosonic and fermionic fields obey periodic and anti-periodic boundary conditions in the time direction respectively,
\begin{eqnarray}
\phi (x_0+\beta, x_1)= \phi (x_0,x_1),\\
\chi (x_0+\beta, x_1)=-\chi (x_0,x_1).
\end{eqnarray}
While doing the decoupling transformation one has to take into account the appropriate Jacobian.
\begin{eqnarray}
 \mathcal{D}\psi\mathcal{D}\bar{\psi}= J_F  \mathcal{D}\chi\mathcal{D}\bar{\chi},\\
 \mathcal{D}A_\mu= J_A \mathcal{D} \phi  \mathcal{D} \eta.
\end{eqnarray}
 The first Jacobian $J_F$ is not trivial
due to the anomaly and the fact that we
perform a chiral transformation. It is computed for 
finite temperature in Fujikawa method\cite{fuji} and is given by
\begin{eqnarray}
J_F=e^{-\frac{1}{2\pi}\int d^2 x (\partial_\mu \phi)^2}.
\end{eqnarray}
And the bosonic Jacobian is  is given by,
\begin{eqnarray}
J_A= \det(-\frac{\nabla^2}{g}),
\end{eqnarray}
and working on a Lorentz Gauge we find out\cite{naonprd,naon},
\begin{eqnarray}
Z_S=N_0 N_\beta\int \mathcal{D}\phi \mathcal{D}\chi\mathcal{D}\bar{\chi} e^{-\int d^2x \mathcal{L}_{eff}},
\end{eqnarray}
where,
\begin{eqnarray}
\mathcal{L}_{eff}=-i\bar{\chi} \slashed{\partial}\chi +\frac{1}{2e^2} \phi \Box\Box \phi - \frac{1}{2\pi} \phi \Box \phi +im_0\bar{\chi} e^{2 \gamma_5 \phi } \chi.
\end{eqnarray}
Now defining  the finite temperature bosonic propagator ${\Delta'_{F}}^{T}(x)$,
\begin{eqnarray}
[\frac{1}{e^2}\Box\Box-\frac{1}{\pi}\Box]{\Delta'_{F}}^{T}(x) =\delta^2(x).
\end{eqnarray}
Solving,
\begin{eqnarray}
\Delta'_F{^T}(x)=\pi ( \Delta_F^T (0,x)-\Delta_F^T (\frac{e}{\sqrt{\pi}},x)),
\end{eqnarray}
here, $\Delta_F^T (m_0,x)$
denotes a free  scalar propagator with mass $m_0$ at finite temperature in two dimensions. And it can be written as\cite{chupa},
\begin{eqnarray}
\Delta_F^T (m_0,x)=\frac{1}{2\pi}K_0(m_0\beta |Q(x)|),
\end{eqnarray}
whereas,
\begin{eqnarray}
\Delta_F^T (0,x)=-\ln(\frac{\beta c |Q(x)|}{\sqrt{\pi}}),
\end{eqnarray}
where c is a numerical constant (related to Euler's
constant).
Here, the  dimensionless “generalized coordinates” 
$Q\equiv (Q_0, Q_1)$ are\cite{chupa},
\begin{eqnarray}
Q_0=-\cosh(\frac{x_1\pi}{\beta})
\sin(\frac{x_0\pi}{\beta})\\
Q_1=-\sinh(\frac{x_1\pi}{\beta})
\cos(\frac{x_0\pi}{\beta})
\end{eqnarray}
And the massless fermionic propagator at finite temperature in terms of generalized coordinates is,
\begin{eqnarray}
S(x)=\frac{i}{\beta}\frac{\slashed{Q}(x)}{Q^2(x)}
\end{eqnarray}
Now the partition  function can be written as,
\begin{eqnarray}
Z_S=CZ_{BE}
\underbrace{
\sum_{n=0} ^\infty \frac{(-im_0)^n}{n!} \langle \prod_{j=1} ^n\int d^2x_j  \bar{\chi}(x_j)e^{2 \gamma_5 \phi } \chi(x_j)\rangle_{_T}}_\text{contribution due to perturbative expansion over mass parameter}.
\end{eqnarray}
Here, 
$\langle .. \rangle_T$ denotes the thermal average over unperturbed ensemble and
$Z_{BE}$ 
is the Bose-Einstein distribution  for massive boson\cite{pathria} of mass $\frac{e}{\sqrt{\pi}}$.
This is the contribution
from the massless  part of the Lagrangian (see appendix to find out why it is equal to $Z_{BE}$). 
 $C$ is an irrelevant
 constant related to the zero-point energies which can be absorbed in normalization constant.
$Z_{BE}$ is defined as\cite{pathria},
\begin{eqnarray}
\ln Z_{BE}=-\int dk \ln{(1-e^{-\beta \sqrt{k^2+\frac{e^2}{\pi}}})},
\end{eqnarray}
In order to compute the partition function, one has to separate the boson
factor from the free fermionic part by using the identity,
\begin{eqnarray}
\bar{\chi}e^{2\gamma_5\phi}\chi= e^{2\phi}\bar{\chi}\frac{1+\gamma_5}{2}\chi + 
e^{-2\phi}\bar{\chi}\frac{1-\gamma_5}{2}\chi.
\end{eqnarray}
Therefore, the partition function now becomes (ignoring constant zero point energy), 
\begin{eqnarray}
Z_S= Z_{BE}\sum_{k=0} ^\infty \frac{(-im_0)^{2k}}{k!^2} \prod_{j=1} ^{k}\int d^2x_j d^2 y_j <I_1>_{boson}
<I_2>_{fermion},
\end{eqnarray}
where $I_1$ and $I_2$ are defined as,
\begin{eqnarray}
&&I_1=e^{2\sum_i(\phi(x_j)-\phi(y_j))}\\
&&I_2=\prod_{j=1} ^k   \bar{\chi}(x_j) \frac{1+\gamma_5}{2}\chi(x_j )
 \bar{\chi}(y_j) \frac{1-\gamma_5}{2}
 \chi(y_j),
\end{eqnarray}
Now using this well-known identity,
\begin{eqnarray}
\langle e^{-i\sum_j\nu_j \phi(x_j)  }\rangle_{bosonic}&=&e^{-\frac{1}{2} \sum_{i,j}\nu_i \nu_j[ -\Delta'^T_F(x_i-x_j)]}, 
\end{eqnarray}
we can solve the bosonic part of the thermal average and using the well known properties of $\gamma_5$ we can rewrite eq. (28) as,
\begin{eqnarray}
I_2= \prod_{j=1}^k \bar{\chi}_1(x_j)\chi_1(x_j)
\bar{\chi}_2(y_j)\chi_2(y_j),
\end{eqnarray}
where, 
\begin{eqnarray}
\chi=
\begin{bmatrix}
    \chi_1       \\
    \chi_2       
\end{bmatrix}.
\end{eqnarray}
This quantity $\prod_{j=1}^k  <\bar{\chi}_1(x_j)\chi_1(x_j)
\bar{\chi}_2(y_j)\chi_2(y_j)>$, i.e. the fermionic part of thermal average   can be easily calculated
for massless fermions\cite{chupa},
\begin{eqnarray}
\prod_{j=1}^k  \langle\bar{\chi}_1(x_j)\chi_1(x_j)
\bar{\chi}_2(y_j)\chi_2(y_j)\rangle=(-1)^k \frac{
\prod_{i>j}^k(
\beta|Q(x_i-x_j|)^2
(\beta|Q(y_i-y_j|)^2
}{\prod_{i,j}^k(\beta|Q(x_i-y_j|)^2}.
\end{eqnarray}
As a result, the partition function becomes,
\begin{eqnarray}
Z_S &=& Z_{BE} \sum_{k=0} ^\infty (\frac{(\frac{mec}{\sqrt{\pi}})^{2k}}
{(k!)^2}) \prod_{j=1} ^{k}\int d^2x_j d^2 y_j \underbrace{\frac{\sum_{i>j} |\beta Q(x_i-x_j)|^2  |\beta  Q(y_i-y_j)|^2}{\sum_{i,j}|\beta Q(x_i-y_j)|^2}}_\text{massless free fermion contribution}
\underbrace{\frac{\sum_{i>j} |\beta Q(x_i-x_j)|^{-2}  |\beta  Q(y_i-y_j)|^{-2}}{\sum_{i,j}|\beta Q(x_i-y_j)|^{-2}}}_\text{massless bosonic contribution}
\nonumber\\
&&
\underbrace{\exp \{ -2 \sum_{i>j} K_0 (\frac{\beta e}{\sqrt{\pi}} |Q(x_i-x_j)|) + K_0 (\frac{\beta e}{\sqrt{\pi}} |Q(y_i-y_j)|)-K_0 (\frac{\beta e}{\sqrt{\pi}} |Q(x_i-y_j)|)\}}_\text{massive bosonic contribution}
\nonumber \\
 &=&Z_{BE}\sum_{k=0} ^\infty \frac{(\frac{mec}{\sqrt{\pi}})^{2k}}
{(k!)^2}
 \prod_{j=1} ^{k}\int d^2x_j d^2 y_j \exp \{ -2 \sum_{i>j} [K_0 (\frac{\beta e}{\sqrt{\pi}} |Q(x_i-x_j)|) + K_0 (\frac{\beta e}{\sqrt{\pi}} |Q(y_i-y_j)|)\nonumber\\
&&-K_0 (\frac{\beta e}{\sqrt{\pi}} |Q(x_i-y_j)|)]\}.
\end{eqnarray}
Here, $m=\frac{m_0}{2\pi}$.
We can see that the massless bosonic excitation  cancels out the fermionic contribution to
$Z_S$.
Turning our attention towards the Lagrangian of the massive sine-Gordon model,
\begin{eqnarray}
\mathcal{L}_{SG}=\frac{1}{2}(\partial_\mu \phi)^2+\frac{1}{2}m_{_{SG}}\phi^2-\frac{\alpha}{\lambda^2}cos(\lambda \phi)
\end{eqnarray}
The partition function for this model,
\begin{eqnarray}
Z_{SG}=N_0 N_\beta \int  \mathcal{D}\phi e^{-\int d^2x \mathcal{L}_{SG}}.
\end{eqnarray}
where the integration runs over scalar fields is periodic
in the time direction,
\begin{eqnarray}
\phi(x_0+\beta,x_1)=\phi(x_0,x_1)
\end{eqnarray}
With the help of bosonic identity eq. (29), 
the partition function of massive sine Gordon model (after identifying $m_{_{SG}}=\frac{e}{\sqrt{\pi}}$),
\begin{eqnarray}
Z_{SG} =&&Z_{BE}\sum_{k=0} ^\infty (\frac{\alpha}{\lambda^2})^{2k}
 (\frac{1}{k!})^2
 \prod_{j=1} ^{k}\int d^2x_j d^2 y_j \exp \{ -\frac{\lambda^2}{2\pi}\sum_{i>j} K_0 (\frac{\beta e}{\sqrt{\pi}} |Q(x_i-x_j)|) + K_0 (\frac{\beta e}{\sqrt{\pi}} |Q(y_i-y_j)|)\nonumber\\
&&-K_0 (\frac{\beta e}{\sqrt{\pi}} |Q(x_i-y_j)|)\}
\end{eqnarray}
Comparing $Z_{S}$ and $Z_{SG}$, we see that the two partition functions are identical
provided the relations,
\begin{eqnarray}
&&\frac{\alpha}{\lambda^2}=\frac{mec}{\sqrt{\pi}}\\
&&\lambda^2=4\pi
\end{eqnarray}
These are the exact duality transformation for the zero temperature case studied by Naon\cite{naonprd}.
 The perturbative series in the
mass term of the massive Schwinger model is found to be term-by-term identical with the perturbative
series in $\alpha$ for the massive sine-Gordon model, provided the relations among the parameters of two models given
in Eqs. (38)-(39) are taken into account. 
Finally, we have proved using the path integral framework that  compactification
of the time variable into a circle with radius $\beta = 1/T$ conserve the
equivalence between the massive sine-Gordon and the massive Schwinger models at fixed radius  (i.e. fixed temperature) $\beta$, just like $T=0$.
We can go into several directions 
based on the result of this work.
For instance,
following the recent development of calculating entanglement entropy at zero temperature vacuum
for Sine Gordon\cite{new1} and Schwinger \cite{new2} models,
 it would be intriguing to  calculate the entanglement entropy of these models at finite temperature, given the duality is now established at finite temperature. Besides that, It has been already proved that this duality holds even for curved space background\cite{curved} at zero temperature. We would like to check  the status of the
 duality  for curved space at finite temperature.
\\\\
\textit{\textbf{Acknowledgments}}\\\\
I would like to express my deep gratitutde to
 Maxim Emelin,
Keshav Dasgupta, Jim Cline, 
 Simon Caron-Hout, Matt Hodel, Shalu Solomon and  Onirban Islam for their valuable
time to discuss several aspects of this topic. 
Many thanks to Kharusi Bursary award for the financial support. 
Finally, I am
very 
thankful for the kind hospitality of
Institute for Advanced study, Princeton where the report is completed.
\section{Appendix: Finite temparature spectrum of Massless Schwinger model}
The Lagrangian density
of the massless Schwinger model in Euclidean (1+1)-dimensional space-time is given as,
\begin{eqnarray}
\mathcal{L}=-\bar{\psi} (i\slashed{\partial}+eA)\psi -\frac{1}{4}F^{\mu \nu}F_{\mu\nu} .
\end{eqnarray}
The partition function can be written as,
\begin{eqnarray}
Z=N_0 N_\beta\int \mathcal{D}A_\mu \mathcal{D}\psi\mathcal{D}\bar{\psi} e^{-\int_\beta d^2x \mathcal{L}_s},
\end{eqnarray}
where, $\int_\beta d^2 x =\int_{0}^{\beta}  dx_0 \int_{-\infty}^{\infty} dx_1$ with, $\beta=\frac{1}{T}$.
Also, the functional integral is performed over fermionic fields satisfying
antiperiodic boundary conditions in the time direction, 
\begin{eqnarray}
\psi(x_0,x_1)=-\psi(x_0+\beta,x_1).
\end{eqnarray}
Now, according to ref. \cite{love}, the partition function at finite temperature can be written as,
\begin{eqnarray}
ln Z= 2\int \frac{dk}{2\pi}\{\frac{k\beta}{2}+ln(1+e^{-\beta k})\} + \int \frac{dk}{2\pi}\{\frac{\beta}{2} (k-k')
+ln \frac{1-e^{-k\beta}}{1-e^{-k'\beta}}\},
\end{eqnarray}
where,
$k'^2=k^2+\frac{e^2}{\pi}$.
It has been quoted in reference\cite{naon} \footnote{see text after equation 3.27} that the finite temperature  partition
function of massless Schwinger model is not equal to  corresponding to the massive boson times free massless fermions, as one could have naively expected. But rather in this partition function 
there is also a factor associated to the zero-mass gauge excitation which
appears in the Lowenstein-Swieca solution for the massless Schwinger model.
This is in disagreement  with the result of 
Love\cite{love}. Because according to ref. \cite{love} the finite temperature  particle content
of the theory contains only a massive non interacting boson of mass $\frac{e}{\sqrt{\pi}}$. 
They showed it by simply taking 
the thermal
ensemble average of the Schwinger model Hamiltonian.
And 
it is then shown
to be equivalent to an ensemble of neutral, massive,
noninteracting Bose particles with all massless
excitations absent. This is clearly not the conclusion of ref. \cite{naon}. But final form of the partition function  of ref.\cite{naon} (i.e. eq. 42) actually  agrees secretly with Love\cite{love}. It can be verified in some very simple steps. Rewriting eq. (42),
\begin{eqnarray}
ln Z= ln Z_{vac} +
2 ln Z_F
- ln Z_{BE}
+ln Z_{BE} ',
\end{eqnarray}
where, 
\begin{eqnarray}
&&\ln Z_{vac}=\int \frac{d k}{2\pi}(\frac{3 }{2} k-\frac{1}{2}k')\beta,\\
&&\ln Z_F=\int \frac{d k}{2\pi} \ln(1+e^{-\beta k}),\\
&&\ln Z_{BE}=-\int \frac{d k}{2\pi} \ln(1-e^{-\beta k}),\\
&&\ln Z_{BE}'=-\int \frac{d k}{2\pi} \ln(1-e^{-\beta k'}).
\end{eqnarray}
Here, $\ln Z_{vac}$, $\ln Z_{F}$, $\ln Z_{BE}$, $\ln Z_{BE}'$ denote vacuum energy, free massless fermionic, free massless bosonic and free massive  bosonic contribution
(with mass $\frac{e}{\sqrt{\pi}}$)
to the the partition function.
Now using the following integral formulas\cite{pathria, mathur}\footnote{See for instance eq. (3.42) and (3.43) of ref.\cite{mathur}},
\begin{eqnarray}
-\int \frac{d k}{2\pi} ln(1-e^{-\beta k})=\frac{1}{\beta}\frac{\pi^2}{6},\\
2\int \frac{d k}{2\pi} ln(1+e^{-\beta k})=\frac{1}{\beta} \frac{\pi^2}{6},
\end{eqnarray}
we can rewrite the partition function.
Due to the existence of minus sign in front of $ln Z_B$ in eq. (6),
massless fermionic and massless bosonic contributions cancel each other and we are left with,
\begin{eqnarray}
\ln Z= ln Z_{vac} +
\ln Z_B '.
\end{eqnarray}
Using the usual definition, $\langle H\rangle=-\frac{\partial}{\partial \beta}lnZ$ we easily find out from eq. (6),
\begin{eqnarray}
\langle H \rangle_T= \langle H \rangle_{T=0}+\int\frac{dk}{2\pi}\frac{\sqrt{k^2+\frac{e^2}{\pi}}}{e^{\sqrt{k^2+\frac{e^2}{\pi}}}-1},
\end{eqnarray}
which is the final result of Love (eq. 45 of ref. \cite{love}).
Therefore, it can now be concluded that finite temperature particle spectrum in both of the approaches match with each other, predicting 
 finite temperature Schwinger model  contains only a massive non interacting boson of mass $\frac{e}{\sqrt{\pi}}$.\\

\end{document}